\documentclass[]{spie}  
\usepackage[]{graphicx}
\usepackage{multirow}
\usepackage{amsmath}
\usepackage{url}
\usepackage{bm}
\def\bicep{{\sc Bicep}}
\def\bicepone{{\sc Bicep1}}
\def\biceptwo{{\sc Bicep2}}
\def\bicepthree{{\sc Bicep3}}

\def\keckarray{{\it Keck Array}}
\def\planck{{\it Planck}}
\def\wmap{{\sc Wmap}}
\def\emode{$E$-mode}
\def\bmode{$B$-mode}

\def\healpix{{\tt Healpix}}
\def\spider{{\sc Spider}}

\title{Optical characterization of the BICEP3 CMB polarimeter at the South Pole}

\author{K.~S.~Karkare\supit{a}, 
P.~A.~R.~Ade\supit{b},
Z.~Ahmed\supit{c,d},
K.~D.~Alexander\supit{a},
M.~Amiri\supit{e},
D.~Barkats\supit{a},
S.~J.~Benton\supit{f,g},
C.~A.~Bischoff\supit{a},
J.~J.~Bock\supit{h,i},
H.~Boenish\supit{a},
R.~Bowens-Rubin\supit{a},
I.~Buder\supit{a},
E.~Bullock\supit{j},
V.~Buza\supit{a},
J.~Connors\supit{a},
J.~P.~Filippini\supit{h,k,l},
S.~T.~Fliescher\supit{m},
J.~A.~Grayson\supit{c,d},
M.~Halpern\supit{e},
S.~A.~Harrison\supit{a},
G.~C.~Hilton\supit{n},
V.~V.~Hristov\supit{h},
H.~Hui\supit{h},
K.~D.~Irwin\supit{c,d,n},
J.~H.~Kang\supit{c,d},
E.~Karpel\supit{c},
S.~Kefeli\supit{h},
S.~A.~Kernasovskiy\supit{c},
J.~M.~Kovac\supit{a},
C.~L.~Kuo\supit{c,d},
E.~M.~Leitch\supit{o},
M.~Lueker\supit{h},
K.~G.~Megerian\supit{i},
V.~Monticue\supit{c},
T.~Namikawa\supit{c,d},\\
C.~B.~Netterfield\supit{f,p},
H.~T.~Nguyen\supit{i},
R.~O'Brient\supit{i},
R.~W.~Ogburn IV\supit{c,d},
C.~Pryke\supit{j,m},
C.~D.~Reintsema\supit{n},
S.~Richter\supit{a},
M.~T.~St.~Germaine\supit{a},
R.~Schwarz\supit{m},
C.~D.~Sheehy\supit{o},
Z.~K.~Staniszewski\supit{h,i},
B.~Steinbach\supit{h},
G.~P.~Teply\supit{h},
K.~L.~Thompson\supit{c,d},
J.~E.~Tolan\supit{c,d},
C.~Tucker\supit{b},
A.~D.~Turner\supit{i},
A.~G.~Vieregg\supit{a,o},
A.~Wandui\supit{c,d},
A.~Weber\supit{i},
J.~Willmert\supit{m},
C.~L.~Wong\supit{a},
W.~L.~K.~Wu\supit{q,c,d},
and K.~W.~Yoon\supit{c,d}
\skiplinehalf
\supit{a}Harvard-Smithsonian Center for Astrophysics, Cambridge,
MA 02138, USA \\
\supit{b}School of Physics and Astronomy, Cardiff University, Cardiff,
CF24 3AA, United Kingdom \\
\supit{c}Department of Physics, Stanford University, Stanford,
CA 94305, USA \\
\supit{d}Kavli Institute for Particle Astrophysics and Cosmology, SLAC
National Accelerator Laboratory, Menlo Park, CA 94025, USA \\
\supit{e}Department of Physics and Astronomy, University of British
Columbia, \\ Vancouver, BC, V6T 1Z1, Canada \\
\supit{f}Department of Physics, University of Toronto, Toronto,
ON, M5S 1A7, Canada \\
\supit{g}Department of Physics, Princeton University, Princeton,
NJ 08544, USA \\
\supit{h}Department of Physics, California Institute of Technology,
Pasadena, CA 91125, USA \\
\supit{i}Jet Propulsion Laboratory, Pasadena, CA 91109, USA \\
\supit{j}Minnesota Institute for Astrophysics, University of
Minnesota, Minneapolis, MN 55455, USA \\
\supit{k}Department of Physics, University of Illinois at 
Urbana-Champaign, Urbana, IL 61801, USA\\
\supit{l}Department of Astronomy, University of Illinois at
Urbana-Champaign, \\ Urbana, IL 61801, USA\\
\supit{m}School of Physics and Astronomy, University of Minnesota, Minneapolis,
MN 55455, USA \\
\supit{n}National Institute of Standards and Technology, Boulder,
CO 80305, USA \\
\supit{o}University of Chicago, Chicago, IL 60637, USA \\
\supit{p}Canadian Institute for Advanced Research, Toronto, ON,
M5G 1Z8, Canada \\
\supit{q}Department of Physics, University of California, 
Berkeley, CA 94720, USA
}

\authorinfo{Send correspondence to K.~S.~Karkare: 60 Garden St., MS
  42, Cambridge, MA 02138, USA.  E-mail: kkarkare@cfa.harvard.edu\\}

\begin{document} 
  \maketitle 

\begin{abstract}
\bicepthree\ is a small-aperture refracting cosmic microwave
background (CMB) telescope designed to make sensitive polarization
maps in pursuit of a potential \bmode\ signal from inflationary
gravitational waves.  It is the latest in the
\bicep/\keckarray\ series of CMB experiments located at the South
Pole, which has provided the most stringent constraints on inflation
to date.  For the 2016 observing season, \bicepthree\ was outfitted
with a full suite of 2400 optically coupled detectors operating at 95
GHz.  In these proceedings we report on the far field beam performance
using calibration data taken during the 2015-2016 summer deployment
season \emph{in situ} with a thermal chopped source.  We generate
high-fidelity per-detector beam maps, show the array-averaged beam
profile, and characterize the differential beam response between
co-located, orthogonally polarized detectors which contributes to the
leading instrumental systematic in pair differencing experiments.  We
find that the levels of differential pointing, beamwidth, and
ellipticity are similar to or lower than those measured for
\biceptwo\ and \keckarray.  The magnitude and distribution of
\bicepthree 's differential beam mismatch -- and the level to which
temperature-to-polarization leakage may be marginalized over or
subtracted in analysis -- will inform the design of next-generation
CMB experiments with many thousands of detectors.

\end{abstract}



\section{INTRODUCTION}
\label{sec:intro}  

A period of accelerated exponential expansion in the early Universe,
while a radical extrapolation from understood physics, naturally
solves the flatness, horizon, and monopole problems of standard
cosmology\cite{planck_p15xx_15}.  This inflationary period also
explains the origin of structure by stretching quantum fluctuations to
macroscopic scales.  Inflation predicts perturbations to the metric in
both scalars (density waves) and tensors (gravitational waves).  Both
types of perturbations affect the polarization of the cosmic microwave
background (CMB) at last scattering\cite{polnarev85}: scalars can only
generate a gradient-type (\emode) pattern, while tensors can also
generate a curl-type (\bmode) pattern.  Measurement of \bmode\ power
in the CMB at degree angular scales in excess of the expectation from
gravitational lensing would be direct evidence for an inflationary
period, and the amplitude of the signal -- parametrized by $r$, the
tensor-to-scalar ratio -- would indicate the energy scale of
inflation.

Since 2006, the \bicep/\keckarray\ CMB polarization experiments have
been mapping $\sim 1\%$ of low-foreground sky from the Amundsen-Scott
South Pole Station.  The telescopes are optimized to detect
fluctuations at degree angular scales, and therefore use a compact,
on-axis refracting optical system with the minimum aperture necessary
to resolve the $\ell \sim 100$ peak in the inflationary gravitational
wave \bmode\ spectrum.  In 2014 we reported a detection of
\bmode\ power at degree angular scales at 150 GHz by
\biceptwo\cite{biceptwo14_I}, which was quickly confirmed by data from
the \keckarray\cite{biceptwo15_V}.  Multifrequency maps are required
to separate Galactic foreground emission (thermal dust at high
frequencies and synchrotron at low frequencies) from the CMB signal.
Using all available \planck\ and \wmap\ maps and the latest
\keckarray\ maps at both 95 and 150 GHz, we find that the excess at
150 GHz is consistent with dust emission, and have constrained the
tensor-to-scalar ratio to $r < 0.07$ at $95\%$
confidence\cite{biceptwo15_BKP_all,biceptwo16_VI}.  To push this limit
further -- or achieve a detection of primordial \bmode s -- we have
deployed 220 GHz detectors in the \keckarray\ to measure dust at high
precision, and have deployed \bicepthree\cite{wu15} at 95 GHz to probe
deeply at a low-foreground frequency.  For the 2016 observing season,
\bicepthree\ was outfitted with a full complement of 20 detector tiles
(2400 optically coupled detectors).

The \bicep/\keckarray\ telescopes measure polarization by differencing
co-located, orthogonally polarized detector pairs\cite{biceptwo14_II}.
Any mismatch in beam shape between the two polarizations will leak
some of the bright CMB temperature signal into polarization.
Temperature-to-polarization leakage is the most prominent instrumental
systematic effect which may cause a false \bmode\ signal and must be
precisely characterized\cite{biceptwo15_IV}.  We have developed an
analysis technique called deprojection\cite{biceptwo15_III} to
marginalize over any false polarization signal which would arise from
the coupling of the CMB temperature sky to a second-order expansion of
the measured differential beam; however, the contribution from
higher-order beam mismatch remains.  In the \biceptwo\ final results,
using high-fidelity far field beam maps we constrained the
undeprojected residual contribution to the final \bmode\ spectrum to
an equivalent $r < 3.0 \times 10^{-3}$.

In these proceedings, we present an analysis of the far field beam
response of the \bicepthree\ detectors observing in the 2016 season.
In February and March of 2016, we obtained far field beam maps of all
\bicepthree\ detectors \emph{in situ} at the South Pole using a
chopped thermal source in the far field of the telescope.  These
measurements serve several purposes: the array-averaged beam profile
is used to smooth \healpix\ maps for simulations, the measured beam
shapes offer a cross-check that the modes removed by deprojection
correspond to actual beam mismatch and also allow for direct
subtraction of the deprojection templates, and the beam maps are used
in dedicated simulations to explicitly predict the residual
\bmode\ power after deprojection.

In Section~\ref{sect:optics} we review the optical design of
\bicepthree.  In Section~\ref{sect:ffbm} we describe the far field
beam measurement setup and the new chopped source constructed in 2016.
In Section~\ref{sect:beamparam} we present two-dimensional Gaussian
fits to each detector and characterize beamwidth, ellipticity,
differential pointing, differential beamwidth, and differential
ellipticity across the entire array.  In Section~\ref{sect:composites}
we coadd all beam measurements to obtain high signal-to-noise
composite maps of individual beams, and use these beams to calculate
the array-averaged beam response.  In Section~\ref{sect:deprojection}
we discuss the measured beam mismatch in the context of deprojection
and the residual beam systematics levels achieved by \biceptwo\ and
\keckarray.  Three companion papers -- a \bicepthree\ status
update\cite{grayson16}, \bicepthree\ detector performance\cite{hui16},
and measurements of \keckarray\ detector polarization angles with a
dielectric sheet calibrator\cite{bullock16} -- are also presented at
this conference.

\section{OPTICAL DESIGN}
\label{sect:optics}

\begin{figure}[h!]
\begin{center}
\includegraphics[width=0.9\columnwidth]{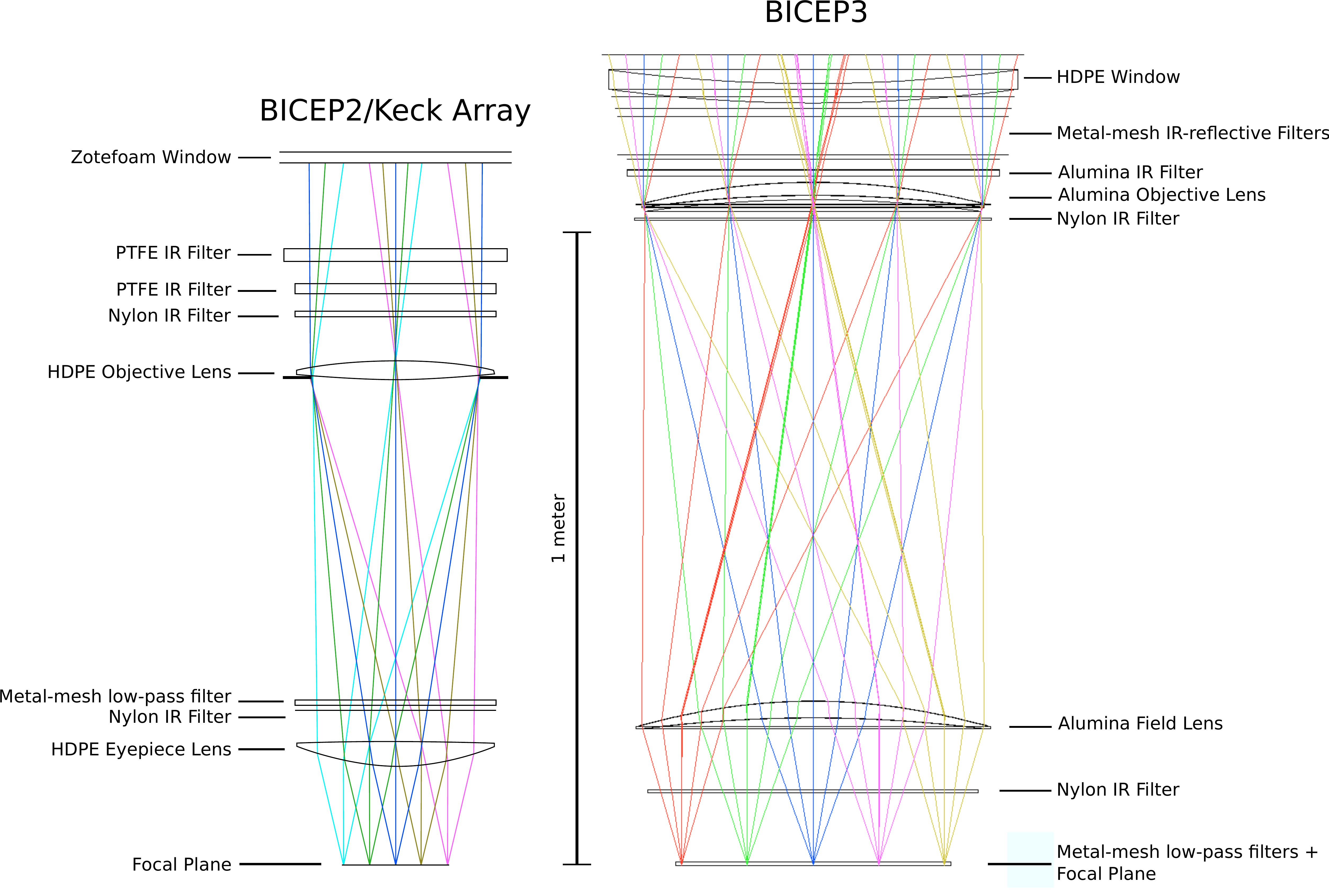}
\caption{\label{fig:optics} Optical schematic of
  \biceptwo/\keckarray\ (left) and \bicepthree\ (right) with optical
  elements labeled, to scale.  \bicepthree's larger aperture (520 mm
  vs 264 mm) and faster optics result in optical throughput similar to
  that of five \keckarray\ receivers combined.}
\end{center}
\end{figure}

Like \bicepone, \biceptwo\, and \keckarray, \bicepthree\ is a compact,
two-lens on-axis telecentric refracting telescope\cite{ahmed14}.
However, the aperture is twice as large as that of its predecessors
($520$ mm vs $264$ mm), the field of view is larger ($27^{\circ}$ vs
$15^{\circ}$), and the optics are faster ($f/1.7$ vs $f/2.4$).  For
observations at a wavelength of $\lambda = 3 \mbox{ mm}$, the far
field is at roughly $2D^2 / \lambda = 180 \mbox{ m}$.

Figure~\ref{fig:optics} shows a schematic of the optical system
compared to that of \biceptwo/\keckarray.  Mounted directly on the
detector modules are metal-mesh low pass filters with a cutoff at $4
\mbox{ cm}^{-1}$ (120 GHz) to block out-of-band power.  Two lenses,
made of 99.6\% pure alumina ceramic, perform the imaging; a cold Lyot
stop skyward of the objective lens defines the aperture.  The
telescope housing, lenses, and aperture stop are all cooled to 4 K.
The alumina lenses are anti-reflection coated with a mix of Stycast
1090 and 2850 epoxy tuned to the correct index of refraction ($n=1.75$
for alumina at $n=3.1$), machined to the correct thickness and laser
diced into patches to relieve mechanical stress from differential
thermal contraction.

The $670 \mbox{ mm}$ aperture high-density polyethylene vacuum window
is $31.75 \mbox{ mm}$ thick and anti-reflection coated with Teadit
24RGD.  The large aperture allows over $100 \mbox{ W}$ of infrared
power to enter the cryostat, and the purely absorptive filtering used
in \biceptwo\ and \keckarray\ would overload the Cryomech PT415 pulse
tube; we therefore reject a majority of the loading with a stack of 10
metal-mesh IR-reflective filters mounted just below the receiver
window at ambient temperature.  Below these are flat absorptive
filters: a $10 \mbox{ mm}$ thick alumina filter at 50 K above the
objective lens, a $5 \mbox{ mm}$ thick nylon filter at 4 K above the
field lens, and $9.5 \mbox{ mm}$ thick nylon filter at 4 K between the
field lens and focal plane.  There is an additional reflective metal
mesh filter below the alumina filter at 50 K.  The resulting IR load
allows continuous operation of the pulse tube at $< 4$ K and $> 48$
hour hold time of the sub-Kelvin refrigerator.

To terminate far sidelobes in CMB observations, we install a co-moving
absorptive forebaffle coated with a combination of Eccosorb
HR-25/AN-75 and weatherproofed with Volara foam.  The forebaffle is
$128.9 \mbox{ cm}$ tall, $141 \mbox{ cm}$ in diameter, and intercepts
radiation $> 14^{\circ}$ from the telescope boresight.  The telescope
is surrounded by a stationary reflective ground shield which redirects
off-axis rays to cold sky; the geometry requires that a ray must
diffract twice -- around the forebaffle and ground shield -- before it
hits the ground.

\section{FAR FIELD BEAM MEASUREMENT SETUP} 
\label{sect:ffbm}

The far field beam mapping setup for \bicepthree\ is very similar to
that used for \biceptwo\ and the \keckarray; see the Beams
paper\cite{biceptwo15_IV} for additional details.

\subsection{Flat Mirror and Thermal Chopped Microwave Source} 
\label{ss:mirrorchopper}

\begin{figure}[h!]
\begin{center}
\includegraphics[width=1.0\columnwidth]{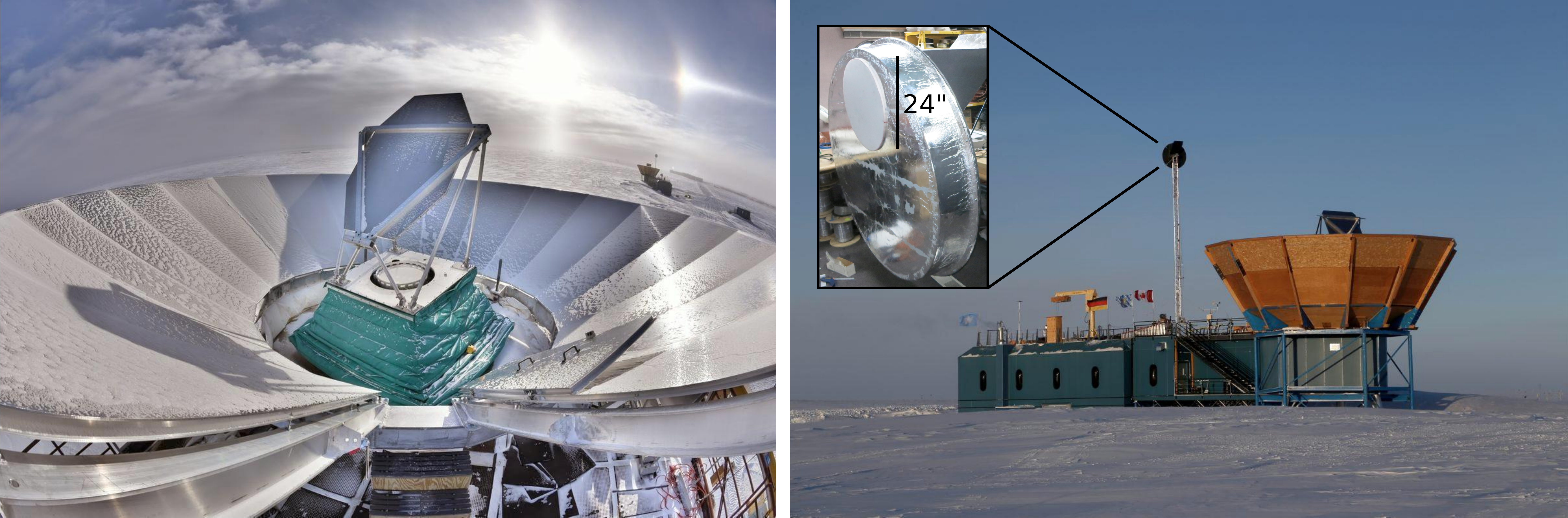}
\caption{\label{fig:mirrorchopper} Left: The aluminum honeycomb mirror
  mounted on \bicepthree\ at the Dark Sector Laboratory in the far
  field beam mapping configuration.  The source is visible in the
  distance.  Right: The chopped thermal source mounted on a 40 ft tall
  mast at the Martin A. Pomerantz Observatory.  The \keckarray\ flat
  mirror is also visible.  Right inset: the chopped source in lab.
  The 24'' aperture is sealed with white HD-30 Zotefoam.  The
  reflective housing contains a blackened rotating blade, behind which
  is a flat mirror redirecting up to sky.}
\end{center}
\end{figure}

Since the telescope is contained within its ground shield and can only
tip down to $\sim 40^{\circ}$ from zenith, we use a flat mirror to
redirect the beams over the ground shield and towards the thermal
chopper (Figure~\ref{fig:mirrorchopper}, left panel).  The 1.7 x 2.5 m
aluminum honeycomb mirror is flat to 0.2 mm across its surface and is
mounted at a $45^{\circ}$ angle, such that when the telescope is
pointed at zenith, the beams are redirected towards the horizon (the
source is $\sim 2^{\circ}$ above the horizon).  The forebaffle is
removed to provide room for the mirror.

For the 2016 deployment season, we built two identical chopped thermal
microwave sources so that \bicepthree\ and \keckarray\ could take far
field beam map data simultaneously.  These choppers consist of a
rotating blade, an enclosure surrounding the blade, and a motor drive
assembly to spin the blade.  The blade is a composite material made
from carbon fiber, low density foam, and honeycomb Nomex, coated with
Eccosorb HR-10.  The enclosure is made of a similar carbon fiber
composite, coated with aluminum foil so that rays which do not
terminate on the source aperture are reflected to cold sky (see the
right inset of Figure~\ref{fig:mirrorchopper}).  The source aperture
is 24'' in diameter, behind which is a flat mirror directed to zenith,
so that the source chops between ambient ($\sim 260$ K) and sky ($\sim
10$ K).  The entire assembly weighs 55 pounds.  The right panel of
Figure~\ref{fig:mirrorchopper} shows the chopper mounted on a 40 ft
high mast on the Martin A. Pomerantz Observatory, 211 meters away from
\bicepthree.  The chopper can spin at a maximum of 9 Hz (18 Hz chop
rate due to the two blades); for beam mapping in 2016 we operated at a
14 Hz chop rate.  The aperture is sealed with HD-30
Zotefoam\footnote{HD-30 is also used for the \keckarray\ vacuum window
  and has $< 2\%$ transmission loss in our band.} to prevent air
turbulence from slowing down the blade.  Compared to previous chopped
thermal microwave sources deployed at South Pole, the 24'' diameter
chopper offers higher signal due to the larger aperture and lower
noise due to the faster chop rate.

\subsection{2016 Data Set} 
\label{ss:dataset}

The far field beam maps presented in these proceedings were taken in
February and March of 2016.  Due to the comparatively high loading of
the $\sim 250$ K source, beam maps are taken with the detectors on the
aluminum superconducting transition\footnote{Our transition edge
  sensor (TES) detectors contain superconductors with different
  transition temperatures suitable for various loading conditions:
  titanium for CMB observations and aluminum for the higher loading in
  calibration measurements.}.  In this higher-loading configuration
the sub-Kelvin refrigerator lasts for 12 hours.  We scan across the
source in $30^{\circ}$ azimuth rasters at $0.8^{\circ}$ per second,
increasing the elevation by $0.05^{\circ}$ in between each scan.  The
scan speed was chosen so that at the fast detector sample rate (150
Hz) and source chop rate (14 Hz), multiple chop cycles would be
recorded across the $0.167^{\circ}$ beam.  The full field of view of
the instrument is therefore covered in $\sim 8$ hours.  To fill in the
gap before the next fridge cycle, we additionally run ``half''
schedules which only cover the top, middle, or bottom of the focal
plane.  Schedules are evenly divided among five boresight angles:
$0^{\circ}$, $45^{\circ}$, $90^{\circ}$, $135^{\circ}$, and
$315^{\circ}$.  Consistency of beam maps across boresight angles
offers a robust check on the repeatability of our beam measurements.
Including both full and half focal plane configurations, we took a
total of 77 beam mapping schedules.  Once the data are taken, we
demodulate the timestream data using a reference signal from the
chopper and bin into maps with $0.1^{\circ}$ pixelization.  The
reflection off the mirror is handled with a pointing model that
describes both the mount and the mirror.

\section{GAUSSIAN BEAM PARAMETERS} 
\label{sect:beamparam}

\subsection{Coordinate System and 2D Gaussian Fits}
\label{ss:coord}

\begin{figure}[h!]
\begin{center}
\includegraphics[width=0.75\columnwidth]{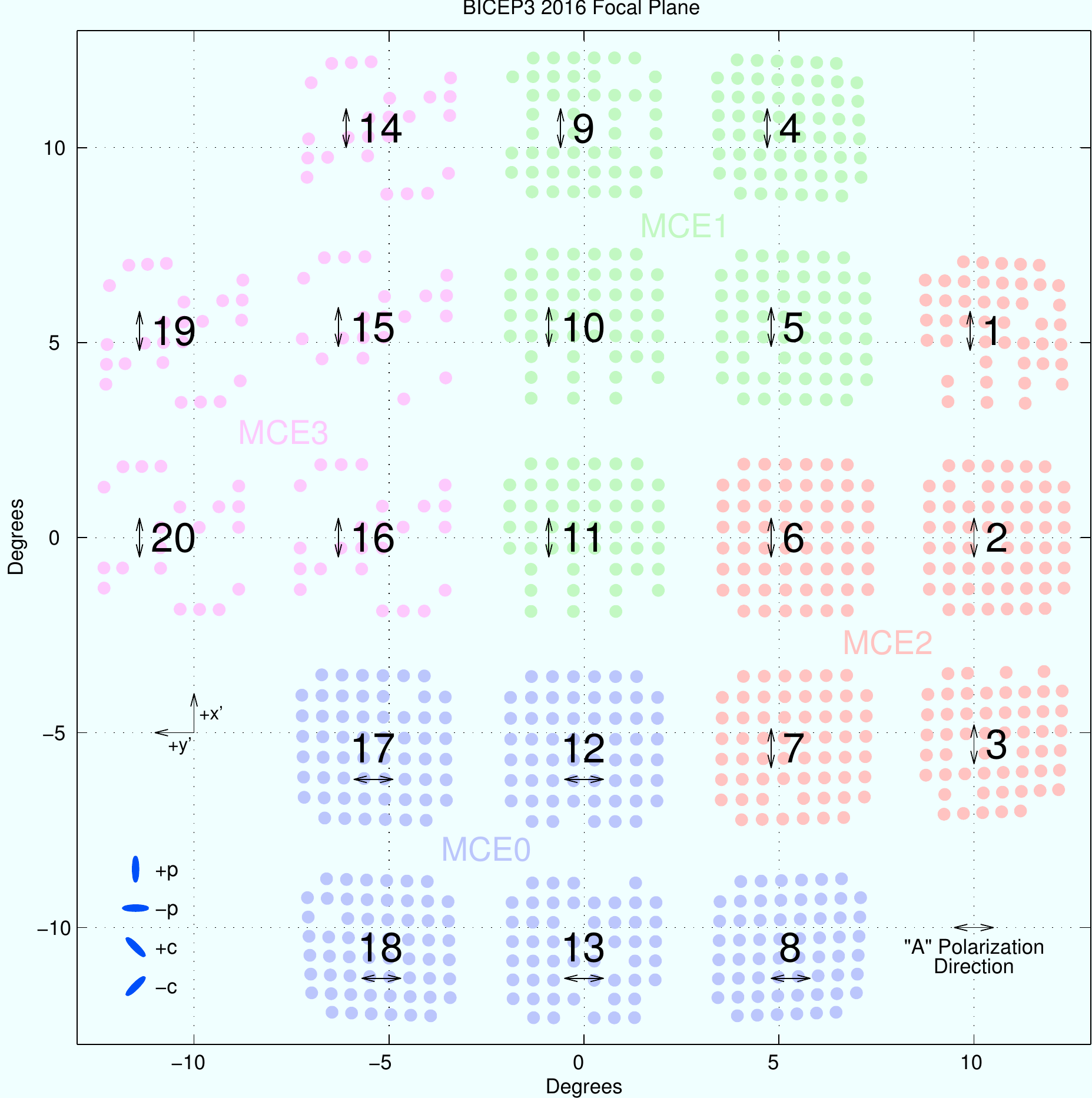}
\caption{\label{fig:fpu} Instantaneous projection of \bicepthree\ beam
  centers on the sky; non-functioning detectors are not plotted.  Tile
  numbers are indicated in black, and the $A$ detector polarization
  direction is indicated with arrows on each tile (the $B$ orientation
  is perpendicular to $A$).  The tiles corresponding to each MCE are
  plotted by color.  The $(x',y')$ coordinate system is indicated, in
  addition to the directions corresponding to plus ($p$) and cross
  ($c$) ellipticity.  Many MCE3 detectors are non-functioning due to
  an electrical open in 7 multiplexing rows.  Tile 21 does not exist
  because more detectors would require another MCE system.}
\end{center}
\end{figure}

To facilitate comparison of beam parameters across measurements taken
at multiple telescope orientations, we define a coordinate system
similar to that used for \biceptwo\ and \keckarray\ beams analysis --
see the Beams paper\cite{biceptwo15_IV} for further details.  Each
pixel $P$, consisting of two orthogonally polarized detectors, is at a
location $(r,\theta)$ from the boresight.  $r$ is the radial distance
away from the boresight, and $\theta$ is the counterclockwise angle
looking out from the telescope toward the sky from the
$\theta=0^{\circ}$ ray, defined to be the great circle that runs along
tiles 11/10/9 on the focal plane.  For each pixel we then define a
local $(x',y')$ coordinate system.  The positive $x'$ axis is along
the great circle passing through $P$ which is an angle $-\theta$ from
the $\hat{r}$ direction of the pixel, while the $y'$ axis is along the
great circle $+90^{\circ}$ counterclockwise away from the $x'$ axis.
The $(x',y')$ coordinate system is then projected onto a plane at $P$.
This coordinate system is fixed to the instrument and rotates on the
sky with the boresight rotation angle $K$ and the angle with which the
receiver is clocked with respect to the $K=0$ line, or the drum angle
$K'$; for \bicepthree, $K'=1.5^{\circ}$.  In a simplified sense, to
transform from mount coordinates to detector-centered coordinates, we
flip the elevation axis to account for the mirror and rotate by $K +
K'$ to arrive at $(x',y')$.  All focal plane and beam map plots in
these proceedings are rendered as the instantaneous projection of each
beam onto the sky, with the same parity (viewing the projection from
inside the celestial sphere) and orientation (with the $+x'$ axis
pointing up and the $+y'$ axis pointing to the left); this is
consistent with the conventions established in our other publications.
Figure~\ref{fig:fpu} shows the coordinate axes, the location of each
nominally working beam, tile numbers and polarization axes, and the
division of tiles into separate Multi-Channel Electronics (MCE)
systems used to control the multiplexing and readout of the detectors.

Each beam is fit to a two-dimensional elliptical Gaussian model with
six free parameters:

\begin{equation}
 B(\bm{x}) = \frac{1}{\Omega} e^{-\frac{1}{2}(\bm{x}-\bm{\mu})^{T} \Sigma^{-1} (\bm{x}-\bm{\mu})},
\end{equation}
where $\bm{x} = (x,y)$ is the location of the beam center, $\bm{\mu} =
(0,0)$ is the origin, $\Omega$ is the normalization, and $\Sigma$ is
the covariance matrix parametrized as

\begin{equation}
 \Sigma = 
\begin{pmatrix}
\sigma^2 (1 + p) & c \sigma^2 \\
c \sigma^2 & \sigma^2 (1 - p)
\end{pmatrix}.
\end{equation}
$\sigma$ is the beamwidth and $p$ and $c$ are the ellipticities in the
``plus'' and ``cross'' directions respectively, such that an
elliptical Gaussian with a major axis oriented along $x'$ or $y'$
would have a $+p$ or $-p$ ellipticity, and one with a major axis
oriented diagonally would have a $\pm c$ ellipticity (see
Figure~\ref{fig:fpu}).  Parametrization with ($\sigma, p, c$), while
equivalent to the more common ($\sigma_{maj}, \sigma_{min}, \phi$), is
more convenient because differential parameters are simple
differences: $d\sigma = \sigma_A - \sigma_B$, $dx = x_A - x_B$, $dy =
y_A - y_B$, $dp = p_A - p_B$, and $dc = c_A - c_B$, where $A$ and $B$
refer to the orthogonally polarized detectors within a pair.  These
parameters directly scale the deprojection templates used to remove
temperature-to-polarization leakage in analysis.  Note that we do not
use the absolute $(x,y)$ beam locations because of the complicated
mirror model, instead obtaining beam centers from cross-correlation
with the \planck\ CMB temperature maps.

In each of the 77 beam map schedules, we find the best fit $\Omega$,
$\bm{x}$, $\sigma$, $p$, and $c$ for every beam.  Not all beams are
covered in each schedule, and occasionally some detectors are not
working normally even when they are covered.  We then remove maps in
which the fit failed, or in which only one of the $A$ or $B$
polarizations were working, so that the same number of measurements
and detectors are used in per-detector and per-pair parameters.  After
cuts, the average pair has $\sim 30$ good measurements.
Table~\ref{tab:stats} shows summary statatistics derived from these
measurements.  For each detector/pair, we take the median across the
$\sim 30$ measurements as the best estimate of each parameter, and
take half the width of the central 68\% of the distribution of those
measurements as the measurement uncertainty.\footnote{This statistic
  is relatively insensitive to outliers, and would equal $1 \sigma$
  for a Gaussian distribution of measurements.}  The characteristic
uncertainty for an individual measurement -- taken to be the median
measurement uncertainty for all detectors/pairs across the array -- is
shown in the ``Individual Measurement Uncertainty'' column of
Table~\ref{tab:stats}.

To characterize the distribution of parameters, we also find the
median across the focal plane (``FPU Median'' column) and quantify the
variation across the focal plane as half the width of the central 68\%
of the distribution of \emph{best estimates for each detector/pair}
(``FPU Scatter'' column).  Note that the ``FPU Scatter'' measures
the spread of best-estimate parameters across the array and is not a
measurement uncertainty.

\begin{table}[t]
\caption{\bicepthree\ 2016 Beam Parameter Summary Statistics.  See
  Section~\ref{ss:coord} for a discussion of measurement
  uncertainties.}
\label{tab:stats}
\begin{center}
\begin{tabular}{lccc} 
\hline
\hline
\rule[-1ex]{0pt}{3.5ex} Parameter & FPU Median & FPU Scatter & Individual Measurement Uncertainty\\
\hline
Beamwidth $\sigma$ (degrees) & 0.167 & 0.002 & 0.002 \\
Ellipticity plus $p$ (+) & 0.010 & 0.021 & 0.026 \\
Ellipticity cross $c$ ($\times$) & -0.004 & 0.016 & 0.026 \\
Diff. X pointing $dx$ (arcmin) & 0.03 & 0.13 & 0.05 \\
Diff. Y pointing $dy$ (arcmin) & -0.12 & 0.17 & 0.05 \\
Diff. Beamwidth $d\sigma$ (degrees) & -0.001 & 0.001 & 0.001 \\
Diff. Ellipticity plus $dp$ (+) & -0.006 & 0.017 & 0.004 \\
Diff. Ellipticity cross $dc$ ($\times$) & 0.000 & 0.005 & 0.004 \\
\hline
\end{tabular}
\end{center}
\end{table}

The median Gaussian beamwidth $\sigma$ across the focal plane is
$0.167^{\circ}$, corresponding to a FWHM of 23.6 arcminutes, and has
been corrected for the non-negligible size of the source aperture (see
Section~\ref{ss:arrayaverage}).  The per-detector ellipticity has
large variation across the focal plane, with a central value close to
zero for both $p$ and $c$.  The ellipticity individual measurement
uncertainty is somewhat large (but similar to those obtained for
\biceptwo/\keckarray) because random artifacts in the beam maps
occasionally cause the beam fitting routine to choose an ellipticity
which is a poor fit to the beam.  These measurements can be improved
by careful deglitching or visually inspecting beam residuals and
removing failed fits.  We also note that the typical measurement
uncertainties of differential parameters are much smaller; this is
because artifacts tend to affect both detectors in a pair equally and
thus bias the parameter estimate in the same way.

\subsection{Per-Pair Parameters} 
\label{ss:perpair}

In pair differencing experiments, differential beam parameters
correspond to components of the intensity field which leak into
polarization.  Here we characterize the distribution of beam
parameters corresponding to the difference of elliptical Gaussians,
which approximately couple to the CMB temperature map and its
derivatives, and which are marginalized over in our deprojection
procedure.  In Section~\ref{sect:deprojection} we compare the
magnitude and distribution of these measurements to those found for
\biceptwo\ and \keckarray.

\begin{figure}[h!]
\begin{center}
\includegraphics[width=0.75\columnwidth]{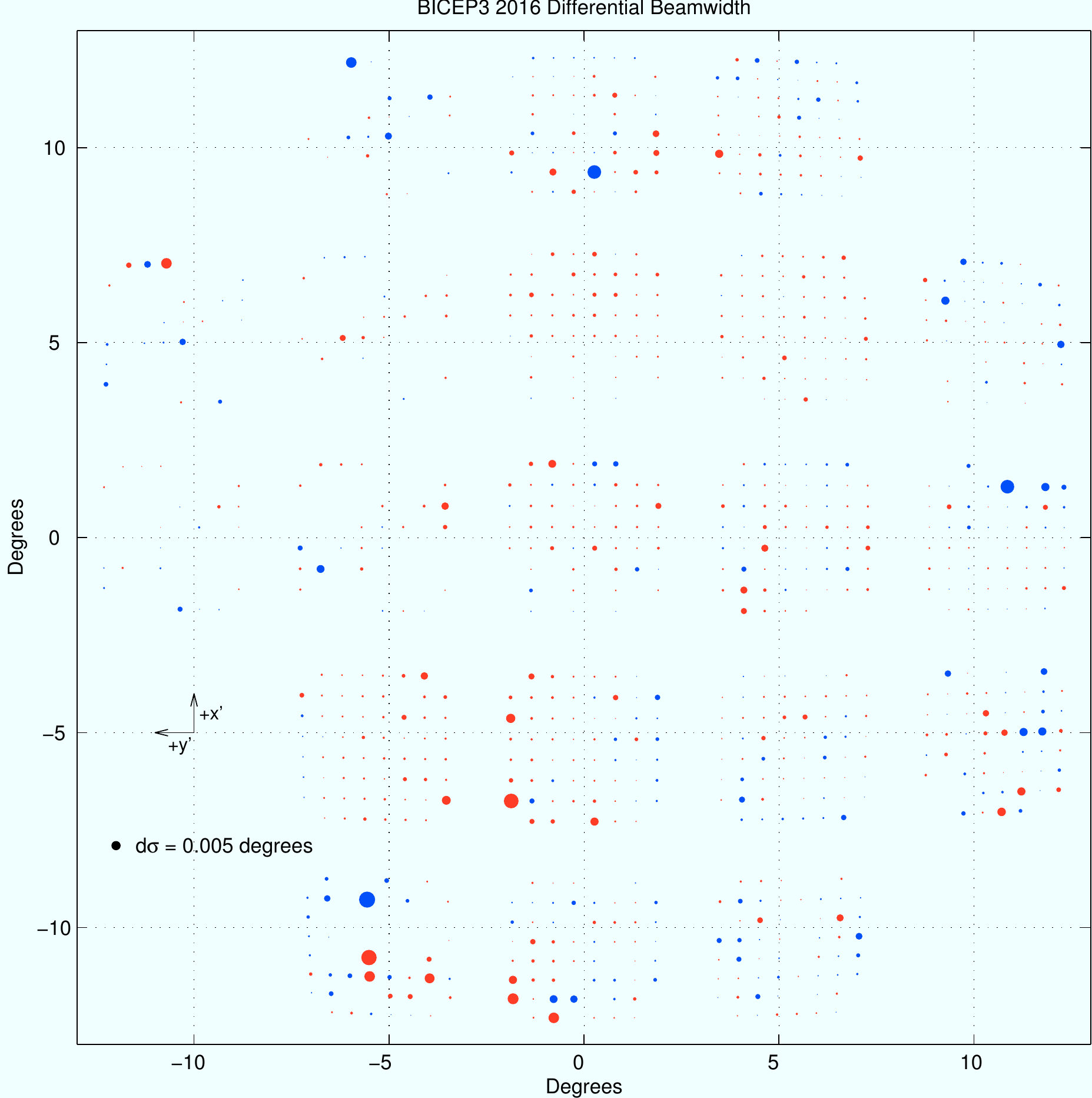}
\caption{\label{fig:diffsig} \bicepthree\ differential beamwidth
  $d\sigma$.  The size of the circle is proportional to the beamwidth
  difference between $A$ and $B$; blue circles indicate $A$ is larger
  than $B$.  The distribution is incoherent across the focal plane and
  small compared to the median beamwidth $\sigma = 0.167^{\circ}$.}
\end{center}
\end{figure}

\begin{figure}[h!]
\begin{center}
\includegraphics[width=0.75\columnwidth]{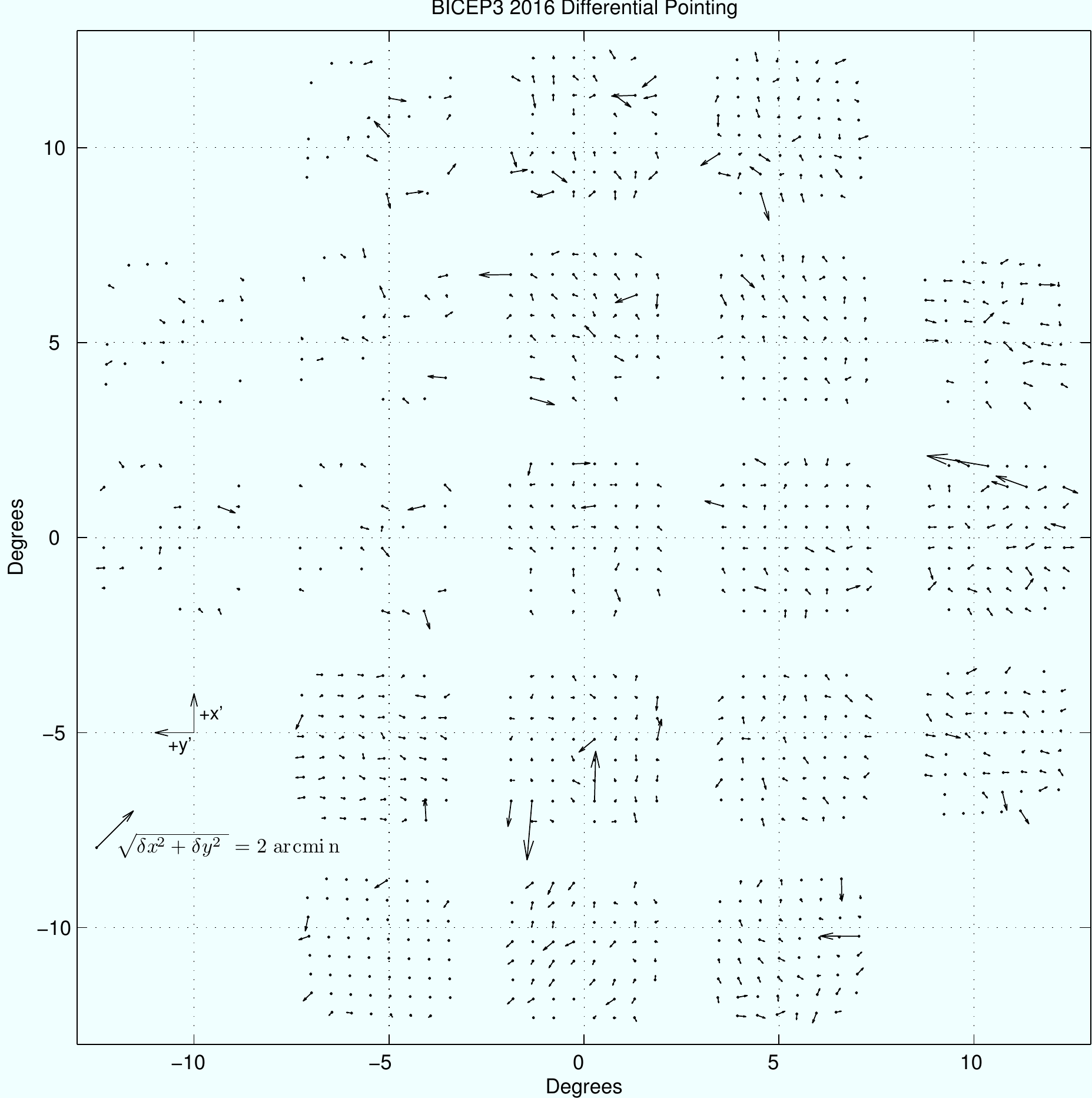}
\caption{\label{fig:diffpoint} \bicepthree\ differential pointing
  $dx$, $dy$.  The arrows point from the $A$ detector location to the
  $B$ detector location, but the tails are centered on the measured
  $A/B$ centroid.  The length of the arrow corresponds to the degree
  of mismatch, scaled up by a factor of 40 so that the small offsets
  are visible.  The distribution does not show a pattern across the
  focal plane, but several tiles have a coherent pointing direction.}
\end{center}
\end{figure}

\begin{figure}[h!]
\begin{center}
\includegraphics[width=0.75\columnwidth]{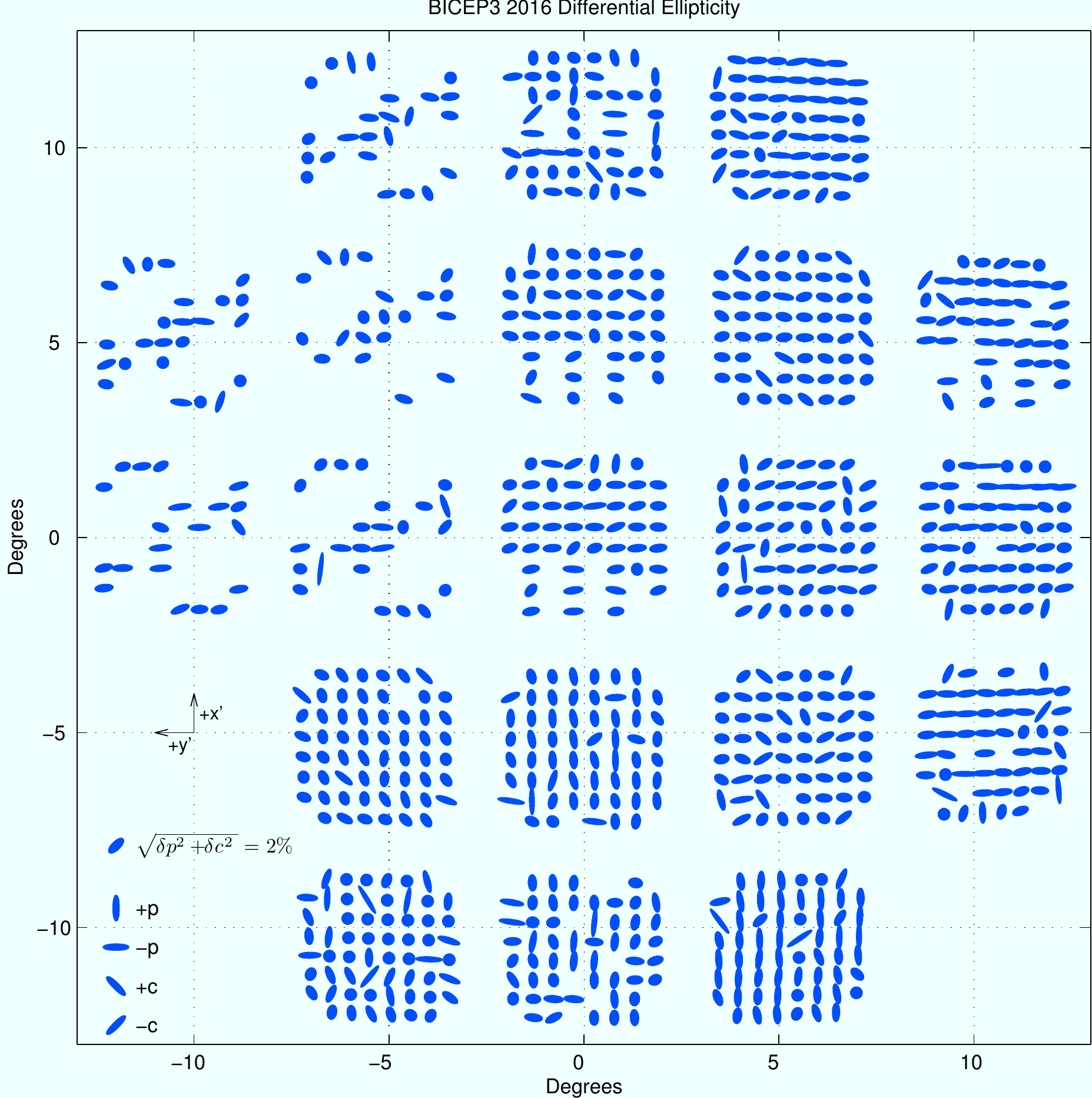}
\caption{\label{fig:diffellip} \bicepthree\ differential ellipticity
  $\sqrt{dp^2 + dc^2}$.  The direction of the ellipse indicates the
  direction of the differential ellipticity -- the legend indicates
  the axes corresponding to the $p$ and $c$ axes.  The ellipticity is
  highly exaggerated to highlight small deviations from the nominal
  circular beams.  Tiles 8, 12, 13, 17, and 18 -- the three in the
  bottom row and the left two in the row immediately above it -- are
  rotated $90^{\circ}$ with respect to the others.  Tiles mostly
  exhibit $\pm dp$ differential ellipticity, which is along the tile
  $B$ polarization axis.}
\end{center}
\end{figure}

Figure~\ref{fig:diffsig} shows the differential beamwidth across the
focal plane.  The distribution is centered around zero, incoherent
across the array, and small compared to the beamwidth -- the median
absolute differential beamwidth is $0.0007^{\circ}$ compared to
$0.167^{\circ}$.  Figure~\ref{fig:diffpoint} shows the differential
pointing, which for \biceptwo\ and \keckarray\ has traditionally been
the largest beam shape mismatch mode.  The distribution appears random
across the focal plane, although several tiles exhibit coherence
across the tile (e.g. Tiles 1 and 17); the median offset magnitude
$\sqrt{dx^2 + dy^2}$ is 0.103 arcminutes.  Figure~\ref{fig:diffellip}
shows the differential ellipticity across the focal plane.  We notice
that differential ellipticity is largely coherent within tiles, and is
mostly $\pm dp$ corresponding to ellipse orientations along the $x'$
or $y'$ axis.  Tiles 8, 12, 13, 17, and 18 (those corresponding to
MCE0; see Figure~\ref{fig:fpu}) are rotated $90^{\circ}$ with respect
to the rest of the tiles.  When their parameters are rotated so that
all tiles have the same polarization orientation (i.e. with $A$ along
the $x'$ axis), we find that differential ellipticity is primarily
$-dp$, corresponding to the $y'$ direction on the focal plane and the
$B$ polarization axis on the tiles.  The median differential
ellipticity for the array in the same orientation is $dp = -0.014$ and
$dc = 0.000$.

\section{COMPOSITE BEAM MAPS AND ARRAY-AVERAGED MAPS} 
\label{sect:composites}

\subsection{Composite Maps}
\label{ss:composites}

\begin{figure}[h!]
\begin{center}
\includegraphics[trim=0 0 0 0,clip,width=1.0\columnwidth]{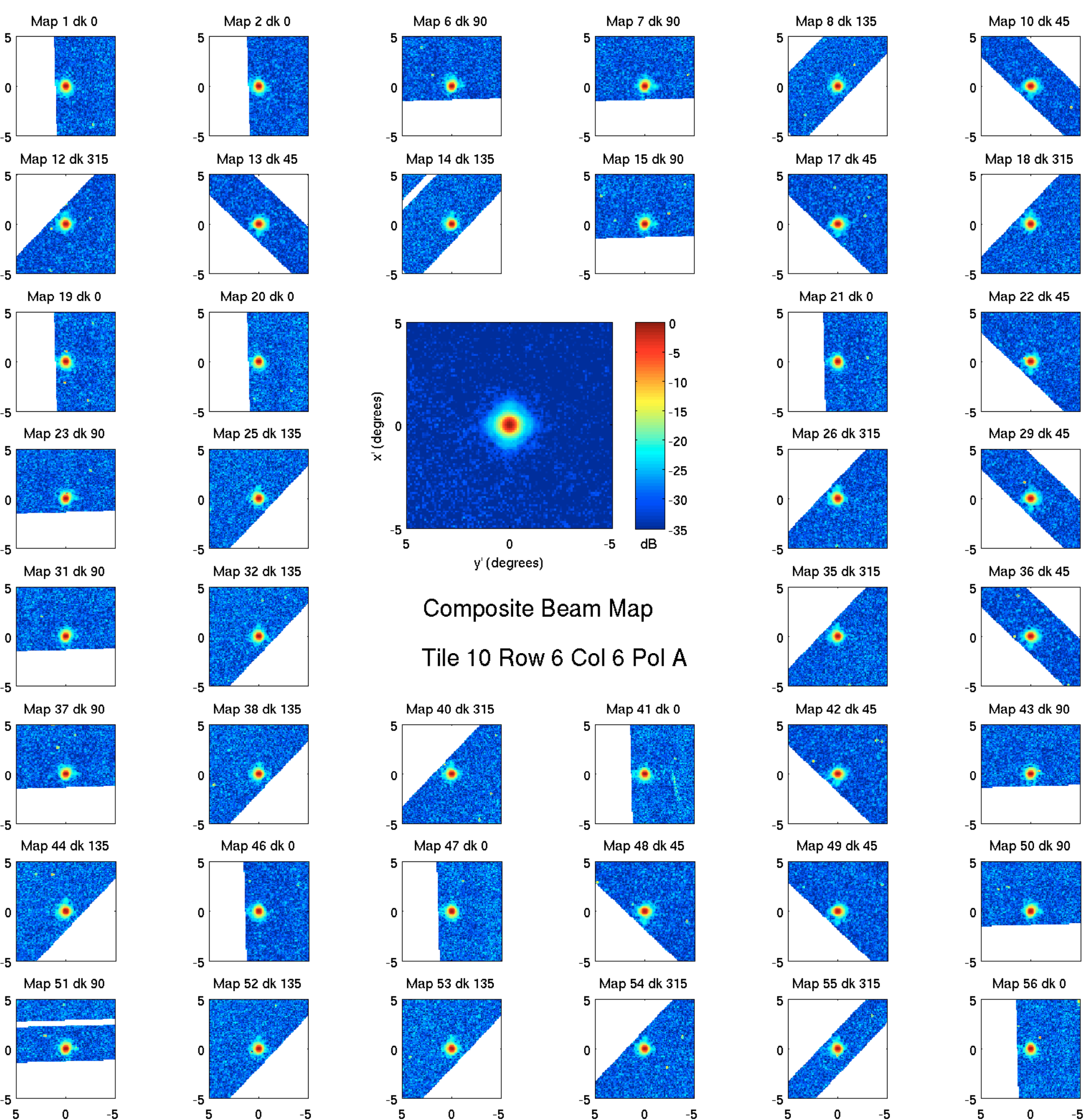}
\caption{\label{fig:composite} Composite beam map for a single
  detector (center), surrounded by the component maps used in the
  composite.  Each component map is labeled with the schedule number
  (out of 77 total) and the boresight rotation angle.  All maps are
  rotated to standard $(x',y')$ coordinates and are in dB relative to
  the peak amplitude -- the composite clearly has a lower noise level
  than the components.  The ground is masked out in each of the
  component maps, $\sim 2^{\circ}$ below the source.}
\end{center}
\end{figure}

Since each detector has many beam maps at multiple boresight rotation
angles, we can coadd them to obtain a high signal-to-noise composite
map which covers all parts of the beam.  We only use beam maps which
have acceptable 2D Gaussian fits and in which both the $A$ and $B$
detectors were measured.  Each component map is rotated to account for
the boresight angle and centered on the common centroid for each
detector pair.  The ground is masked out since it causes visible
response even in the demodulated maps.  The composite map is then made
from the median amplitude of each pixel across component maps.
Figure~\ref{fig:composite} shows an example composite and the
components which were used to make it.  Spurious pixels in the
component maps do not contribute to the final composite, which has
visibly lower noise.  In the 2016 dataset, only 38 optically
responsive detectors have fewer than 10 good component maps (the
typical detector has $\sim 30$).  Although the composite beams shown
in these proceedings only extend to $5^{\circ}$ away from the main
beam, we can generate maps out to much larger angles.  Future work
with extended composite maps will allow for explicit measurement of
extended beam response, including detailed constraints on reflections
and total integrated sidelobe power out to large angles.

\subsection{Array-Averaged Beam}
\label{ss:arrayaverage}

\begin{figure}[h!]
\begin{center}
\includegraphics[width=0.6\columnwidth]{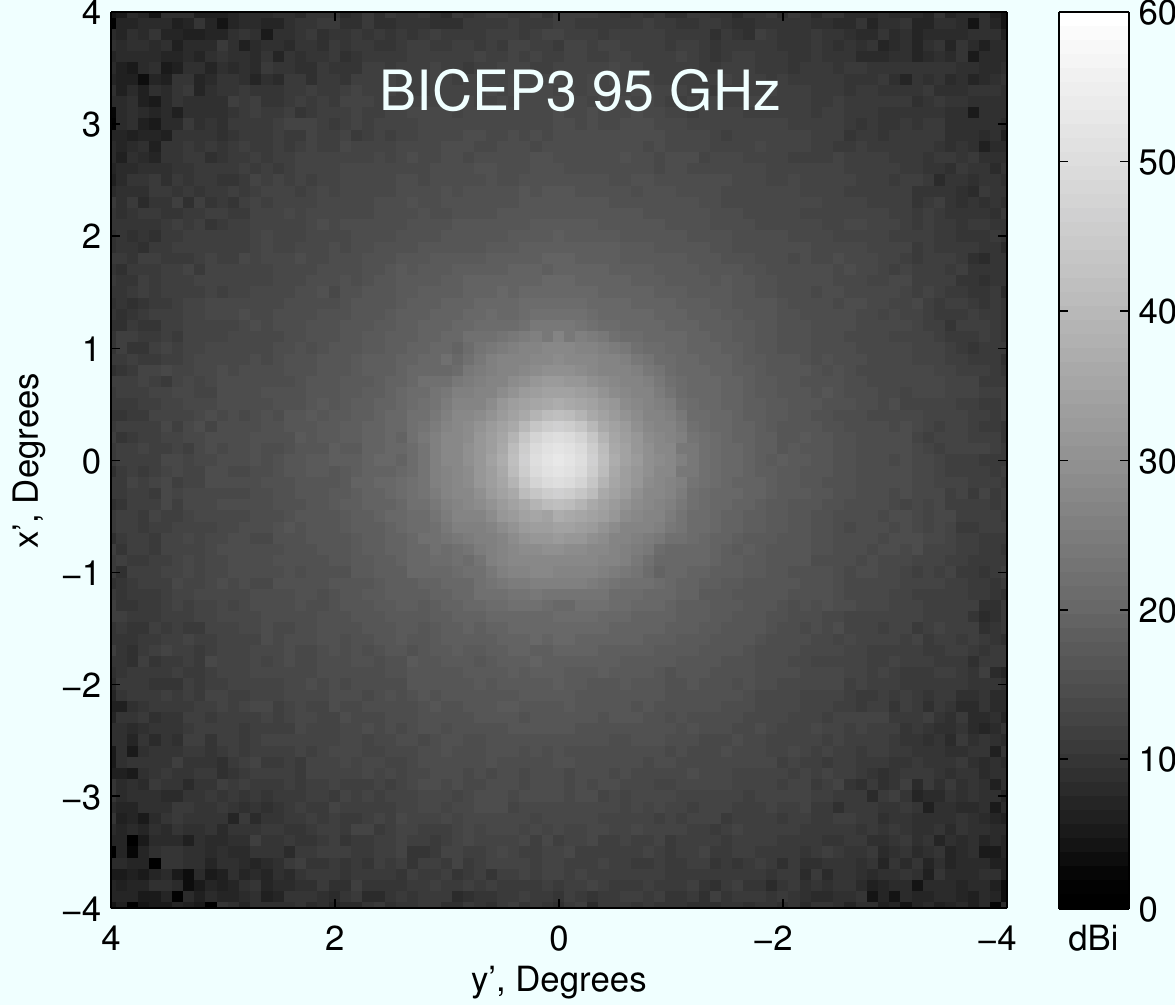}
\caption{\label{fig:avgbeam} \bicepthree\ average beam, made by
  coadding composite beam maps from all optically active detectors.
  The peak is at 52.6 dBi.}
\end{center}
\end{figure}

\begin{figure}[h!]
\begin{center}
\includegraphics[width=0.7\columnwidth]{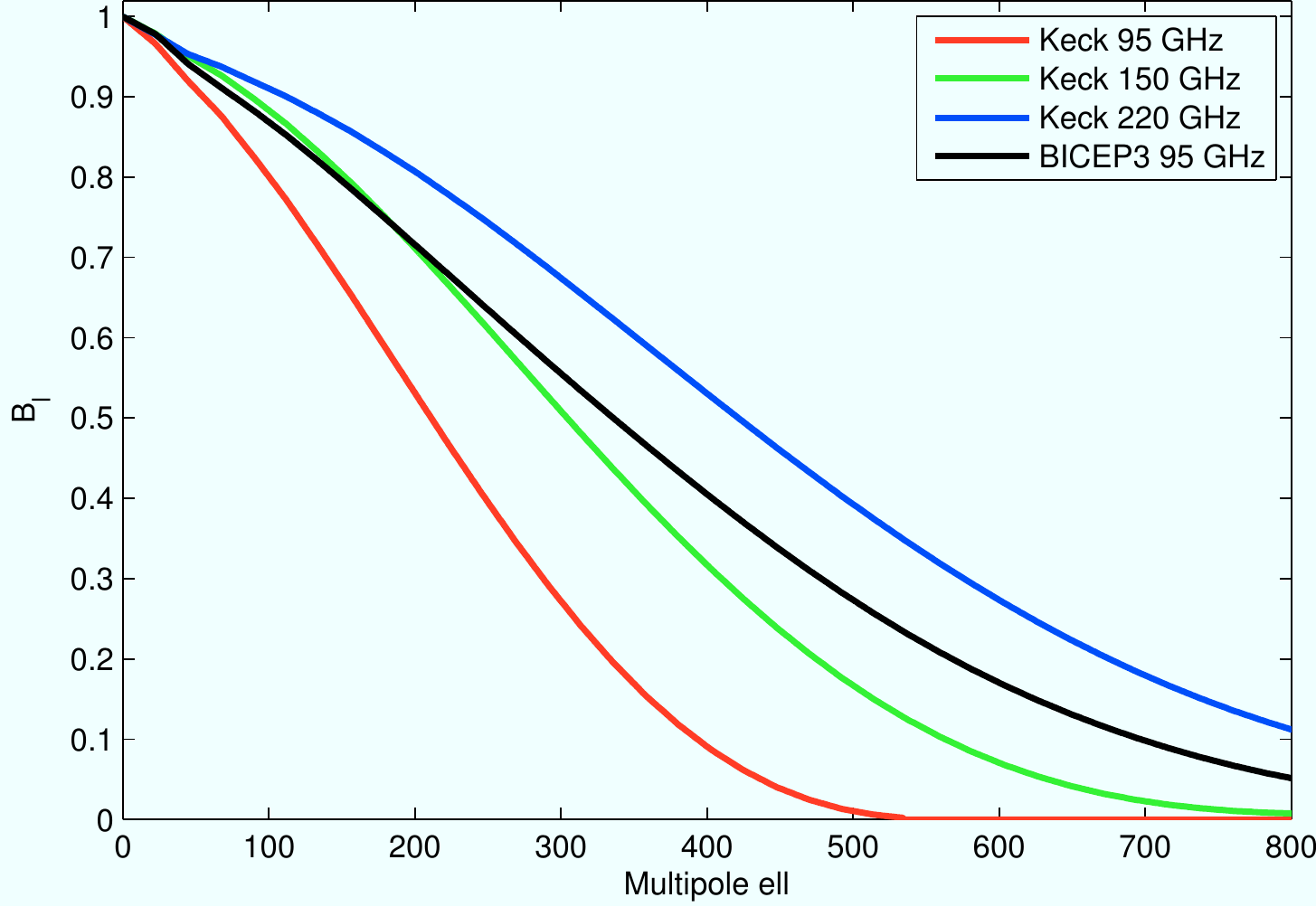}
\caption{\label{fig:bl} \bicepthree\ $B_{\ell}$ profile (black),
  compared to \keckarray\ $B_{\ell}$ at 3 frequencies.  All
  \keckarray\ apertures are 264 mm in diameter, while \bicepthree's is
  520 mm.}
\end{center}
\end{figure}

From the composite beam maps, we integral normalize and coadd all
operational beams to form an array-averaged beam map.
Figure~\ref{fig:avgbeam} shows the averaged map in dBi (i.e. relative
to an isotropic radiator); Airy ring structure is clearly visible.
This map is then averaged in radial bins and Fourier transformed to
obtain the circularly symmetric beam window function $B(\ell)$.  The
measured $B(\ell)$ is corrected for the effect of the finite size of
the source aperture:

\begin{equation}
B(\ell)_{measured} = B(\ell)_{true} \times \frac{2J_1 \left( \ell D/2 \right)}{\ell D/2},
\end{equation}
where $J_1$ is the Bessel function of the first kind and $D =
0.179^{\circ}$ is the diameter of the thermal source as seen from
\bicepthree.  Figure~\ref{fig:bl} shows $B(\ell)_{true}$ for
\bicepthree\ and \keckarray, which we use to smooth the \planck\ input
maps sampled in our timestream-level simulations.

\section{DEPROJECTION AND EXPECTED SYSTEMATICS CONTROL} 
\label{sect:deprojection}

\begin{table}[ht]
\caption{Differential Beam Parameters for \biceptwo, \keckarray, and
  \bicepthree.  Experiments are ordered by decreasing median beam size
  (see Section~\ref{sect:deprojection} for a discussion about the
  scaling of leakage modes with beam size).  Detector count only
  includes optically active detectors -- the total number of TESs is
  higher.  \keckarray\ parameters are combined over all receivers
  operating at the indicated frequency -- five for 150 GHz, and two
  each for 95 and 220 GHz -- so coherent per-receiver effects may
  average out (see e.g. $dx$ in \keckarray\ 2012).  For more detailed
  \keckarray\ beam parameters, see Table 3 of the Beams
  paper\cite{biceptwo15_IV}.  Values are Array Median $\pm$ Array
  Scatter $\pm$ Individual Measurement Uncertainty.}
\label{tab:statsvb2}
\begin{center}
\begin{tabular}{lccc} 
\hline
\hline
\rule[-1ex]{0pt}{3.5ex} Parameter & \keckarray\ 2015 (95 GHz) & \biceptwo\ (150 GHz) & \keckarray\ 2012 (150 GHz) \\
\hline
$\sigma$ (degrees) & 0.306 & 0.220 & 0.216 \\
Detector count & 544 & 500 & 2480 \\
$dx$ (arcmin) & 0.19 $\pm$ 0.51 $\pm$ 0.05 & 0.81 $\pm$ 0.29 $\pm$ 0.14 & 0.00 $\pm$ 0.97 $\pm$ 0.07  \\
$dy$ (arcmin) & -0.24 $\pm$ 0.48 $\pm$ 0.04 & 0.78 $\pm$ 0.35 $\pm$ 0.14 & -0.21 $\pm$ 0.54 $\pm$ 0.06 \\
$d\sigma$ (degrees) & 0.000 $\pm$ 0.001 $\pm$ 0.001 & 0.000 $\pm$ 0.001 $\pm$ 0.001 & 0.000 $\pm$ 0.001 $\pm$ 0.001 \\
$dp$ (+) & -0.010 $\pm$ 0.008 $\pm$ 0.002 & -0.002 $\pm$ 0.013 $\pm$ 0.011 & -0.007 $\pm$ 0.015 $\pm$ 0.003 \\
$dc$ ($\times$) & 0.002 $\pm$ 0.003 $\pm$ 0.002 & -0.003 $\pm$ 0.012 $\pm$ 0.005 & -0.007 $\pm$ 0.009 $\pm$ 0.003 \\
\hline
\end{tabular}

\vspace{10pt}

\begin{tabular}{lccc} 
\hline
\hline
\rule[-1ex]{0pt}{3.5ex} Parameter & \keckarray\ 2013 (150 GHz) & \bicepthree\ (95 GHz) & \keckarray\ 2015 (220 GHz)\\
\hline
$\sigma$ (degrees) & 0.216 & 0.167 & 0.142 \\
Detector count & 2480 & 2400 & 992 \\
$dx$ (arcmin) & 0.11 $\pm$ 0.58 $\pm$ 0.07 & 0.03 $\pm$ 0.13 $\pm$ 0.05  & -0.05 $\pm$ 0.19 $\pm$ 0.04 \\
$dy$ (arcmin) & -0.22 $\pm$ 0.56 $\pm$ 0.06 & -0.12 $\pm$ 0.17 $\pm$ 0.05 &   -0.01 $\pm$ 0.28 $\pm$ 0.04  \\
$d\sigma$ (degrees) & 0.000 $\pm$ 0.001 $\pm$ 0.001 & -0.001 $\pm$ 0.001 $\pm$ 0.001 &  0.000 $\pm$ 0.001 $\pm$ 0.001 \\
$dp$ (+) & -0.011 $\pm$ 0.012 $\pm$ 0.002 & -0.006 $\pm$ 0.017 $\pm$ 0.004 &  -0.017 $\pm$ 0.008 $\pm$ 0.006\\
$dc$ ($\times$) & -0.004 $\pm$ 0.007 $\pm$ 0.002 & 0.000 $\pm$ 0.005 $\pm$ 0.004 & 0.003 $\pm$ 0.007 $\pm$ 0.006 \\
\hline
\end{tabular}

\end{center}
\end{table}

In this section, we compare \bicepthree's differential beam parameters
with those from \biceptwo\ and \keckarray\ (Table~\ref{tab:statsvb2})
and comment on the expected level of temperature-to-polarization
leakage.  In our CMB analysis pipeline, for each detector pair we
perform a deprojection procedure in which the timestream data are
regressed against templates of the real CMB temperature sky and its
derivatives corresponding to the difference modes of two mismatched
beams.  The template multiplied by the best-fit regression coefficient
is removed from the timestream, eliminating all leakage caused by a
particular mode without external knowledge of the beams.  As an
alternative to deprojection, we may also choose to fix the coefficient
which scales the deprojection template to the beam-map-derived value
-- we refer to this procedure as ``subtraction.''  Since differential
gain is likely time-variable, we deproject it.  Gain is not measured
in beam maps because of the different TES used and loading environment
between calibration measurements and CMB observations.  In our
previous results, we have chosen to deproject differential pointing;
the measured correlation between beam-map-derived values and
deprojection coefficients offers a robust check that our beam maps
correspond to reality.  Differential beamwidth has been found to be
negligible, and so has neither been deprojected nor subtracted.  We
subtract differential ellipticity, because the real-sky $TE$
correlation would be filtered out of our $TE$ and $EE$ spectra.

We first consider differential pointing, which for \biceptwo\ and
\keckarray\ has been the beam mismatch mode contributing the most
leakage and is independent of beam size.  While \biceptwo's $dx$ and
$dy$ were quite large, significant development effort\cite{obrient12}
resulted in subsequent \keckarray\ focal planes at 95, 150, and 220
GHz with differential pointing that was several times smaller.
\bicepthree\ is most similar to \keckarray\ 220 GHz in both the
magnitude and the measurement precision.  We can expect \bicepthree's
pre-deprojection leakage from differential pointing to be much less
than that of \biceptwo; the distribution is also incoherent across the
focal plane and will likely average down significantly.

\bicepthree's differential beamwidth is incoherent across the focal
plane, centered close to zero, and is measured with precision similar
to that of previous receivers.  The level of leakage scales with one
power of the beamwidth.  Though in previous results we have found
differential beamwidth to contribute negligible leakage, if beam map
simulations show that this mode is significant for \bicepthree,
deprojection is a possibility.

Differential ellipticity is subtracted in our standard results, and
therefore the precision requirements become stronger.  We first
observe that the measurement uncertainties in Table 2 are lower than
those achieved by \biceptwo, so we may expect that \bicepthree\ will
be able to subtract leakage from differential ellipticity with higher
precision than in previous results.  We also note that the
pre-subtraction leakage scales as the square of the beam size --
\bicepthree's beams are $\sim 75\%$ the size of \biceptwo's, so a
similar level of inherent differential ellipticity would produce
significantly less leakage in \bicepthree.

Ultimately, the effectiveness of deprojection is simulated explicitly
using beam map simulations in which we convolve the \planck\ CMB
temperature sky with the individual beam maps and process simulated
timestreams just like the real data.  We can then deproject these maps
which contain only temperature-to-polarization leakage from measured
beam mismatch.  Based on comparison of the magnitude of the dominant
differential pointing in Table~\ref{tab:statsvb2}, we expect the
inherent leakage from beam mismatch modes to be equal to or lower than
that from \biceptwo.

Though deprojection removes all power corresponding to the coupling of
the CMB temperature sky to a second-order expansion of the measured
differential beam shape, real beams contain higher-order power which
deprojection does not remove.  Since the undeprojected beam mismatch
residuals are often complicated, vary significantly across the focal
plane, and may nontrivially cancel or amplify in the final map, the
degree to which they will affect the final \bmode\ power spectrum is
not immediately obvious.  The undeprojected residual level is the
final result of the beam map simulations after deprojection.  In the
\biceptwo\ final results, we found that after deprojection of relative
gain and differential pointing, and subtraction of differential
ellipticity, we were able to constrain the residual
temperature-to-polarization leakage to an equivalent $r < 3.0 \times
10^{-3}$.  Similar simulations for \bicepthree\ will be run using the
composite beam maps presented above, and we expect the measurement
uncertainty on the beam maps to be much lower than that of
\biceptwo\ because of higher signal strength afforded by the new
thermal chopper.

\section{CONCLUSIONS} 
\label{sect:conclusions}

In these proceedings we have presented a preliminary analysis of
\bicepthree's far field beam performance for the 2016 observing
season.  Measurements were made during the 2015-2016 deployment season
using a new chopped thermal microwave source with a 24'' aperture.  We
find that \bicepthree\ has a median beamwidth of $0.167^{\circ}$ (23.6
arcminutes) and present the array-averaged $B(\ell)$ profile.  When
detector pairs are fit to 2D elliptical Gaussians, we find that
differential pointing, beamwidth, and ellipticity are similar to or
lower than those measured for \biceptwo\ and \keckarray.  The
high-fidelity per-detector composite beam maps we have constructed
from 77 beam mapping schedules will allow for explicit simulation of
the undeprojected residual temperature-to-polarization leakage
expected in the final polarization maps produced by \bicepthree.

\acknowledgments     

The \bicepthree\ project has been made possible through support from
the National Science Foundation (grant Nos. 1313158, 1313010, 1313062,
1313287, 1056465, and 0960243), the SLAC Laboratory Directed Research
and Development Fund, the Canada Foundation for Innovation, and the
British Columbia Development Fund.  The development of antenna-coupled
detector technology was supported by the JPL Research and Technology
Development Fund and grants 06-ARPA206-0040 and 10-SAT10-0017 from the
NASA ARPA and SAT programs.  The development and testing of focal
planes were supported by the Gordon and Betty Moore Foundation at
Caltech.  The computations in these proceedings were run on the
Odyssey cluster supported by the FAS Science Division Research
Computing Group at Harvard University.  Tireless administrative
support was provided by Irene Coyle, Kathy Deniston, Donna Hernandez,
and Dana Volponi.

We are grateful to Samuel Harrison and Hans Boenish as our 2015 and
2016 winterovers, respectively. We thank the staff of the US Antarctic
Program and in particular the South Pole Station without whose help
this research would not have been possible.  We thank our \bicepone,
\biceptwo, \keckarray\, and \spider\ colleagues for useful discussions
and shared expertise.


\bibliography{spie_2016_b3beams}   
\bibliographystyle{spiebib}   

\end{document}